\documentclass
[preprint,byrevtex,pre,nofootinbib,10pt,balancelastpage,titlepage,bibnotes,twocolumn]{revtex4}%
\usepackage{amsfonts}
\usepackage{amsmath}
\usepackage{amssymb}
\usepackage{graphicx}%
\providecommand{\U}[1]{\protect\rule{.1in}{.1in}}
\ifx\pdfoutput\relax\let\pdfoutput=\undefined\fi
\newcount\msipdfoutput
\ifx\pdfoutput\undefined\else
\ifcase\pdfoutput\else
\ifx\paperwidth\undefined\else
\ifdim\paperheight=0pt\relax\else\pdfpageheight\paperheight\fi
\ifdim\paperwidth=0pt\relax\else\pdfpagewidth\paperwidth\fi
\fi\fi\fi
\begin{document}
\preprint{UATP/1703}
\title{Jarzynski Equality and its Special Trajectory Ensemble Average Demystified }
\author{P.D. Gujrati}
\email{pdg@uakron.edu}
\affiliation{Department of Physics, Department of Polymer Science, The University of Akron,
Akron, OH 44325}

\begin{abstract}
The special trajectory ensemble average (TEA) $\left\langle \bullet
\right\rangle _{0}$ in the Jarzynski Equality (JE) results in the Jensen
inequality $\left\langle R\right\rangle _{0}\geq\Delta F$ for the work $R$
done on the system, and not the thermodynamic work inequality $\left\langle
R\right\rangle \geq\Delta F$ since we find $\left\langle R\right\rangle
\neq\left\langle R\right\rangle _{0}$. Therefore, contrary to the common
belief, the Jensen inequality does not directly support the JE as a
nonequilibrium result. Jarzynski's microscopic treatment of the inclusive
energy considers only the external work $d_{\text{e}}E_{k}$ but neglects the
ubiquitous change $d_{\text{i}}E_{k}$ due to external-internal \emph{force
imbalance}, though $d_{\text{i}}E_{k}$'s are present even in a reversible
process as we show. Because of this neglect, no thermodynamic force necessary
for dissipation is allowed. Thus the JE has no built-in irreversibility,
despite a time-dependent work protocol. We support our claim by an explicit
calculation, which shows that $\left\langle R\right\rangle _{0}>\Delta F$ even
for a reversible process for which $\left\langle R\right\rangle =\Delta F$.
This also confirms that $\left\langle R\right\rangle \ $and $\left\langle
R\right\rangle _{0}$ are different averages.

\end{abstract}
\date{November 5, 2017}
\maketitle

\textsc{Introduction }The present work is motivated by a note by Cohen and
Mauzerall (CM) \cite{Cohen} criticizing some aspects of the JE
\cite{Jarzynski} that they found mysterious or disconcerting, and its response
by Jarzynski \cite{Jarzynski-Cohen}. (The use of modern notation
\cite{deGroot,Prigogine,Gujrati-II,Gujrati-Entropy2,Gujrati-Stat} and concepts
summarized in \cite{Note,Note-SI} prove very convenient.) Here, we go beyond
CM and clarify the following mysterious aspects of the JE \cite{Note0},
requiring careful scrutiny they have not received yet:

(i) the nature of the special TEA $\left\langle \bullet\right\rangle _{0}$
used in the JE;

(ii) a common belief that the JE is consistent with the \emph{thermodynamic}
\emph{work inequality }\cite{Note1,Landau} $\left\langle R\right\rangle
\geq\Delta F$ or
\begin{equation}
R_{\text{diss}}=T_{0}\Delta_{\text{i}}S\doteq\left\langle R\right\rangle
-\Delta F\geq0; \label{DissipatedWork}%
\end{equation}
$\left\langle R\right\rangle $ is the \emph{thermodynamic average} (denoted by
$\left\langle \bullet\right\rangle $ above and an overbar $\overline
{\left\langle \bullet\right\rangle }$ by Jarzynski \cite{Jarzynski}) work done
on the system (see Eqs. (\ref{ThermodynamicAverageWork}) and
(\ref{Thermodynamic Average}) for a proper definition) and $R_{\text{diss}}$
the \emph{dissipated work }\cite{Note-R}.

(iii) the JE truly represents a nonequilibrium result.

(iv) a time-dependent work always gives $R_{\text{diss}}>0$.

CM briefly commented on some of them but left them unsettled as their goal was
very different. Our interest is to assess the role and significance of the
TEA's ($\left\langle \bullet\right\rangle _{0}$ and $\left\langle
\bullet\right\rangle \equiv\overline{\left\langle \bullet\right\rangle }$),
and their relationship with the second law, an issue that has not attracted
any interest to the best of our knowledge in view of the belief (ii) above,
but we think is central to a comprehensive and precise understanding of the
JE. Our analysis leads to some unexpected conclusions that not only contradict
commonly held beliefs but also has far reaching consequence wherever
trajectories are exploited. For concreteness, we assume the work process to
change the volume $V(t)$ of the system, but the arguments are valid for any
external work process.

\textsc{The JE }In a remarkable paper \cite{Jarzynski}, Jarzynski used the
fluctuating nature of work $R_{k}$ done on the $k$th microstate $\mathfrak{m}%
_{k}\ $during an arbitrary process$\ \mathcal{P}_{0}$\ to prove the JE%
\begin{equation}
\left\langle e^{-\beta_{0}R}\right\rangle _{0}\doteq%
%TCIMACRO{\tsum \limits_{k}}%
%BeginExpansion
{\textstyle\sum\limits_{k}}
%EndExpansion
p_{k0}e^{-\beta_{0}R_{k}}=e^{-\beta_{0}\Delta F}; \label{JarzynskiRelation0}%
\end{equation}
as $\Delta F$ is an SI quantity \cite{Note-SI}, $R_{k}$ must also be an SI
quantity. Let $\gamma_{k}$ denote the trajectory of $\mathfrak{m}_{k}$ during
$\mathcal{P}_{0}$\ over $(0,\tau_{\text{eq}})$ between two equilibrium states
\textsf{A }and \textsf{B} at the same inverse temperature $\beta_{0}=1/T_{0}$,
$\tau_{\text{eq}}$ being the equilibration time at \textsf{B}, as the system
is driven (the \emph{driving stage}) by a work agent from \textsf{A }to
\textsf{B}. The collection $\left\{  \gamma_{k}\right\}  $ forms the
trajectory ensemble (TE). The work is done during $(0,\tau\leq\tau_{\text{eq}%
})$; we denote the driving stage over $(0,\tau)$ by $\mathcal{P}$ and the rest
by $\overline{\mathcal{P}}$ ($t\geq\tau\leq\tau_{\text{eq}}$). If the system
at $t=\tau$ is out of equilibrium (the state at $t=\tau$ is denoted by
\textsf{b} to distinguish it from its equilibrium counterpart \textsf{B}), it
is brought in thermal contact with a heat source (the \emph{reequilibration
stage}) during $\overline{\mathcal{P}}$ to come to equilibrium at temperature
$T_{0}$. The free energy difference between \textsf{A }and \textsf{B} is
$\Delta F=F_{\mathsf{B}}(\beta_{0})-F_{\mathsf{A}}(\beta_{0})$. Let
$dE_{k}\equiv d_{\text{e}}E_{k}+d_{\text{i}}E_{k}$ be the infinitesimal change
in the energy $E_{k}$ of $\mathfrak{m}_{k}$, where $d_{\text{e}}E_{k}$ and
$d_{\text{i}}E_{k}$ are the changes caused by exchange with the work agent and
due to internal processes, respectively \cite{Note}. Jarzynski does not
recognize the ubiquitous nature of $d_{\text{i}}E_{k}$'s (see below) and
ignores it by setting $dR_{k}(t)=dE_{k}(t)\simeq d_{\text{e}}E_{k}\ $as the
infinitesimal work done on $\mathfrak{m}_{k}$ over $\mathcal{P}$ and
$dR_{k}(t)=0$ over $\overline{\mathcal{P}}$; see, however,
\cite{Note-JE-GeneralizedWork,Gujrati-GeneralizedWork}. Thus, $\Delta
E_{k\mathsf{AB}}\doteq E_{k\mathsf{B}}-E_{k\mathsf{A}}=E_{k\mathsf{b}%
}-E_{k\mathsf{A}}$.\ The \emph{accumulated work }along $\gamma_{k}$ is
$R_{k}\doteq%
%TCIMACRO{\tint }%
%BeginExpansion
{\textstyle\int}
%EndExpansion
dR_{k}(t)$. The TEA $\left\langle \bullet\right\rangle _{0}$ in Eq.
(\ref{JarzynskiRelation0}), to be called the \emph{Jarzynski average},
requires first evaluating $R_{k}$ and then summing the exponential work
$e^{-\beta_{0}R_{k}}$\ over all possible trajectories $\{\gamma_{k}\}$ with
\emph{initial} equilibrium probabilities $p_{k0}=p_{k\mathsf{A}}$ at
\textsf{A}.

The JE is supposed to hold for all (reversible and irreversible) work
processes $\mathcal{P}_{0}$, whether the system remains in thermal contact
with the medium or not during $\mathcal{P}$. For $\mathcal{P}$ to be
irreversible, there must be some "\emph{force imbalance}" between the external
force and the internal force (including friction) it induces as pointed out by
CM \cite{Cohen} and more recently by us \cite{Gujrati-GeneralizedWork}; for
more on this, see later.

\textsc{Ensemble Averages }In general, an equilibrium or nonequilibrium
\emph{ensemble average} (EA) is defined instantaneously, and\ requires
identifying (a): the elements of the ensemble $\left\{  \gamma_{k}\right\}  $
and (b): their instantaneous probabilities $\left\{  p_{k}\right\}  $. The
average is \emph{uniquely} defined over $\left\{  \mathfrak{m}_{k}\right\}  $
using $\left\{  p_{k}\right\}  $\ at each instant, which we identify as the
\emph{instantaneous ensemble average} (IEA). This uniqueness may not hold for
the two TEA's $\left\langle \bullet\right\rangle \equiv\overline{\left\langle
\bullet\right\rangle }$ or $\left\langle \bullet\right\rangle _{0}$. Then care
must be exercised to unravel their physics.

\textsc{Jensen}$^{^{\prime}}$\textsc{s Inequality }By using Jensen's
inequality \textsf{E}$(\Phi(X))\geq\Phi($\textsf{E}$(X))$ for a convex
function $\Phi(X)$ of a random variable $X$, where \textsf{E} is an
expectation operator (such as $\left\langle \bullet\right\rangle
\equiv\overline{\left\langle \bullet\right\rangle }$ or $\left\langle
\bullet\right\rangle _{0}$), Jarzynski \cite{Jarzynski} has argued that the JE
results in $\left\langle R\right\rangle \equiv\overline{R}\geq\Delta F$ in
accordance with Eq. (\ref{DissipatedWork}) for the thermodynamic average work.
The argument is \emph{crucial} since it indirectly justifies the JE to be a
nonequilibrium result. The acceptance of this "proof" is widespread in the
literature and is based on the assumption that the Jarzynski average
$\left\langle R\right\rangle _{0}$ resulting from \textsf{E}$=\left\langle
\bullet\right\rangle _{0}$ is the same as the thermodynamic average
$\left\langle R\right\rangle \equiv\overline{R}$. Indeed, this assumption is
never ever explicitly mentioned but seems to have been accepted by all
workers. To the best of our knowledge, the assumption has never been justified
but is the basis for casting the JE as a nonequilibrium result.

\textsc{Important Results }We first establish that $\left\langle
R\right\rangle _{0}\neq\left\langle R\right\rangle $ by comparing their
definitions. This is a surprising result as it is commonly believed, at least
implicitly, that $\left\langle R\right\rangle _{0}=\left\langle R\right\rangle
$. Thus, the Jensen inequality yields $\left\langle R\right\rangle _{0}%
\geq\Delta F$ that must be distinguished from the thermodynamic work
inequality $\left\langle R\right\rangle \geq\Delta F$. We then carefully
analyze the exclusive (no prime) and inclusive (prime) energy approaches, and
show that they are physically not different by establishing $R_{\text{diss}%
}=R_{\text{diss}}^{\prime}$; however, $\Delta E,\Delta W,\Delta F$, etc. are
different from $\Delta E^{\prime},\Delta W^{\prime},\Delta F^{\prime}$, etc.
but all results from the exclusive approach can be converted to their
inclusive form by simply adding a prime on the quantities involved. Jarzynski
only uses the inclusive approach for which our most important conclusion is
that he neglects the "force imbalance" term that determines $d_{\text{i}}%
E_{k}=-d_{\text{i}}W_{k}$ (see earlier), which is always present, even in a
reversible process as we show. Its absence \emph{necessarily} means that there
is no irreversibility ($R_{\text{diss}}^{\prime}=0$) in the process and
therefore in the JE. To understand the physical significance of $\left\langle
R^{\prime}\right\rangle _{0}\geq\Delta F^{\prime}$, we use an exact
calculation to establish the \emph{strict} inequalities $\left\langle
R\right\rangle _{0}>\Delta F$ and $\left\langle R^{\prime}\right\rangle
_{0}>\Delta F^{\prime}$ for a reversible $\mathcal{P}_{0}$ for which
$R_{\text{diss}}=R_{\text{diss}}^{\prime}=0$. Thus, $\left\langle
R\right\rangle _{0}\geq\Delta F$ or $\left\langle R^{\prime}\right\rangle
_{0}\geq\Delta F^{\prime}$\ as the consequence of the Jensen inequality has
lost its physical significance as a statement of the second law. On the other
hand, we find that the thermodynamic work $\left\langle R\right\rangle =\Delta
F,\left\langle R^{\prime}\right\rangle =\Delta F^{\prime}$ for this reversible
process as expected. We now proceed to establish these results.

\textsc{Tea}'s\textsc{ }In classical thermodynamics \cite{Prigogine,Landau},
the infinitesimal thermodynamic work $\left\langle dR\right\rangle $ done on a
system represents an average quantity. It is used to determine the cumulative
work $\left\langle R\right\rangle $ as an \emph{integral} over the process%
\begin{equation}
\left\langle R\right\rangle \doteq%
%TCIMACRO{\tint \nolimits_{\mathcal{P}}}%
%BeginExpansion
{\textstyle\int\nolimits_{\mathcal{P}}}
%EndExpansion
\left\langle dR\right\rangle . \label{ThermodynamicAverageWork}%
\end{equation}
To get a better understanding of this averaging process, we need to turn to
statistical thermodynamics. The thermodynamic energy is an IEA $\left\langle
E\right\rangle =%
%TCIMACRO{\tsum \nolimits_{k}}%
%BeginExpansion
{\textstyle\sum\nolimits_{k}}
%EndExpansion
p_{k}E_{k}$ over all microstates at a given time $t$. The first law
\cite{Note} during $dt$ is expressed as a sum of two \emph{system-intrinsic}
(SI) contributions \cite{Note-SI}%
\begin{equation}
d\left\langle E\right\rangle =%
%TCIMACRO{\tsum \nolimits_{k}}%
%BeginExpansion
{\textstyle\sum\nolimits_{k}}
%EndExpansion
E_{k}dp_{k}+%
%TCIMACRO{\tsum \nolimits_{k}}%
%BeginExpansion
{\textstyle\sum\nolimits_{k}}
%EndExpansion
p_{k}dE_{k}. \label{FirstLaw}%
\end{equation}
The first sum involving $dp_{k}$ represents the generalized heat $dQ=TdS$
while the second sum involving $dE_{k}$ at fixed $p_{k}$ represents the
\emph{isentropic} change $\left\langle dE\right\rangle $ in the energy
$\left\langle E\right\rangle $ to be identified with $-dW$, the generalized
work \cite{Gujrati-GeneralizedWork,Gujrati-Stat}. It is important to
distinguish $d\left\langle E\right\rangle $ from $\left\langle dE\right\rangle
$ introduced above. It can be written as the IEA $\left\langle dE\right\rangle
\doteq%
%TCIMACRO{\tsum \nolimits_{k}}%
%BeginExpansion
{\textstyle\sum\nolimits_{k}}
%EndExpansion
p_{k}dE_{k}=-%
%TCIMACRO{\tsum \nolimits_{k}}%
%BeginExpansion
{\textstyle\sum\nolimits_{k}}
%EndExpansion
p_{k}dW_{k}=-dW$ in terms of the SI work $dW_{k}=-dE_{k}$ done by the system's
microstate $\mathfrak{m}_{k}$. The cumulative work $\Delta W$ is obtained by
accumulating $dW$ over the entire process $\mathcal{P}_{0}$:%
\begin{equation}
\Delta W\doteq%
%TCIMACRO{\tint \nolimits_{\mathcal{P}_{0}}}%
%BeginExpansion
{\textstyle\int\nolimits_{\mathcal{P}_{0}}}
%EndExpansion
\left\langle dW\right\rangle \doteq%
%TCIMACRO{\tsum \nolimits_{k}}%
%BeginExpansion
{\textstyle\sum\nolimits_{k}}
%EndExpansion%
%TCIMACRO{\tint \nolimits_{\gamma_{k}}}%
%BeginExpansion
{\textstyle\int\nolimits_{\gamma_{k}}}
%EndExpansion
p_{k}(t)dW_{k}(t), \label{ThermodynamicAverageWork0}%
\end{equation}
in which each summand is an integral over the trajectory $\gamma_{k}$, and the
sum is over all trajectories. Each integral involves instantaneous
probabilities $p_{k}(t)$ over the entire trajectory. According to Jarzynski
\cite{Jarzynski}, $dR_{k}=d_{\text{e}}E_{k}$ ($d_{\text{i}}E_{k}=0$ in his
approach; see below) over $\mathcal{P}$ so that $\left\langle dR\right\rangle
$ is given as an IEA similar to $\left\langle dW\right\rangle $. As
$dR_{k}(t)=0$ over $\overline{\mathcal{P}}$, the integration in Eq.
(\ref{ThermodynamicAverageWork}) can be extended to $\mathcal{P}_{0}$:%
\begin{equation}
\left\langle R\right\rangle \doteq%
%TCIMACRO{\tsum \nolimits_{k}}%
%BeginExpansion
{\textstyle\sum\nolimits_{k}}
%EndExpansion%
%TCIMACRO{\tint \nolimits_{\gamma_{k}}}%
%BeginExpansion
{\textstyle\int\nolimits_{\gamma_{k}}}
%EndExpansion
p_{k}(t)dR_{k}(t), \label{Thermodynamic Average}%
\end{equation}
not be confused with $\left\langle R\right\rangle _{0}$, the Jarzynski
average
\begin{equation}
\left\langle R\right\rangle _{0}=%
%TCIMACRO{\tsum \nolimits_{k}}%
%BeginExpansion
{\textstyle\sum\nolimits_{k}}
%EndExpansion
p_{k0}%
%TCIMACRO{\tint \nolimits_{\gamma_{k}}}%
%BeginExpansion
{\textstyle\int\nolimits_{\gamma_{k}}}
%EndExpansion
dR_{k}(t)=%
%TCIMACRO{\tsum \nolimits_{k}}%
%BeginExpansion
{\textstyle\sum\nolimits_{k}}
%EndExpansion
p_{k0}R_{k}. \label{JarzynskiAverage0}%
\end{equation}
We conclude that $\left\langle R\right\rangle _{0}\neq\left\langle
R\right\rangle $ unless $p_{k}(t)=p_{k0},\forall k,t$.

\textsc{Exclusive/Inclusive approaches }We first consider a very common
nonequilibrium example of a gas in a piston to set the stage. The external
pressure on the piston is $P_{0}$, which tries to compress the gas. The gas
responds by adjusting its SI pressure \cite{Note-SI} $P=-\partial E/\partial
V$, which tries to expand the gas. They point in opposite directions and, for
irreversibility, $P\neq P_{0}$ in magnitude. Being an SI quantity, $E$ is a
function of $S$ and $V$, even though we are dealing with a nonequilibrium
state in internal equilibrium \cite{InternalEQ} so $dE=TdS-PdV$ in terms of
the (generalized) heat $dQ=TdS$ and work $dW=PdV$ done by the gas; we do not
consider any internal variables \cite{Gujrati-II} for simplicity. Of $dW$,
$d_{\text{e}}W=P_{0}dV$ (this is the negative of the work $dR=-P_{0}dV$ done
by the external pressure on the gas \cite{Note-R}) is spent to overcome the
external pressure and $d_{\text{i}}W=\Delta PdV,\Delta P\doteq P-P_{0}$, is
the internal work dissipated within the gas \cite{Note}. For the Helmholtz
free energy $H=E+P_{0}V$, we then have $dH=TdS+VdP_{0}-\Delta PdV$
\cite{Note2}. We notice that $H(S,P_{0},V)$ is function of \emph{three}
variables $S,P_{0}$ and $V$ and not \emph{two} ($S,P_{0}$),\ when we are
dealing with a nonequilibrium process. Thus, $H$ is not a Legendre transform
of $E$ with respect to $V$ \cite{Note-Legendre}, unless we deal with a
reversible process ($P=P_{0},d_{\text{i}}W=0$) when $H$ becomes a Legendre
transform $H(S,P_{0})$. Since $P_{0}$ is an external pressure, we can treat it
as a parameter just like $V$. In this case, we can treat $H$ as some new
energy $E^{\prime}(S,P_{0},V)$ with \emph{two} work parameters $P_{0}$ and $V$
\cite{Landau-Mech}. In the terminology of Jarzynski, $E(S,V)$ is an
\emph{exclusive} and $E^{\prime}(S,P_{0},V)$ an \emph{inclusive} energy with
forces $-V=-\partial E^{\prime}/\partial P_{0}\ $and $\Delta P=-\partial
E^{\prime}/\partial V\neq0$ unless $P=P_{0}$. It is convenient to think of the
two energies as SI energies of two different (exclusive and inclusive)
systems, respectively. This allows us to treat both systems in one stoke; all
we need to do is to use a prime on all SI quantities pertaining to the
exclusive system to obtain relations for the inclusive system. The term $TdS$
in both represents the generalized heat $dQ=dQ^{\prime}$ so the generalized
work for the two are: $dW=PdV=P_{0}dV+d_{\text{i}}W$ for $E$ and $dW^{\prime
}=-VdP_{0}+d_{\text{i}}W^{\prime}$ for $E^{\prime}$, where we have introduced
$d_{\text{i}}W^{\prime}=d_{\text{i}}W$ as the dissipated work. We observe that
$dW^{\prime}=-$ $\left(  \partial E^{\prime}/\partial P_{0}\right)  dP_{0}-$
$\left(  \partial E^{\prime}/\partial V\right)  dV$ consists of two work
components due to $P_{0\text{ }}$and $V$, respectively, as suggested above.
This also means that the exchange work is $d_{\text{e}}W=P_{0}dV$ and
$d_{\text{e}}W^{\prime}=-VdP_{0}$. We finally have
\cite{Gujrati-GeneralizedWork}
\[
dW^{\prime}-dW=d_{\text{e}}W^{\prime}-d_{\text{e}}W=-d(P_{0}V),d_{\text{i}%
}W^{\prime}=d_{\text{i}}W.
\]

With the above discussion as a background, we consider the same issue at the
microscopic level. Let $P_{k},P_{k}^{\prime}$ denote the pressures $-\partial
E_{k}/\partial V,-\partial E_{k}^{\prime}/\partial V$ for $\mathfrak{m}_{k}$
in the two approaches. These pressures determine the thermodynamic pressures
$P=%
%TCIMACRO{\tsum \nolimits_{k}}%
%BeginExpansion
{\textstyle\sum\nolimits_{k}}
%EndExpansion
p_{k}P_{k},P^{\prime}=%
%TCIMACRO{\tsum \nolimits_{k}}%
%BeginExpansion
{\textstyle\sum\nolimits_{k}}
%EndExpansion
p_{k}P_{k}^{\prime}$, so that
\begin{equation}
P-P_{0}\doteq%
%TCIMACRO{\tsum \nolimits_{k}}%
%BeginExpansion
{\textstyle\sum\nolimits_{k}}
%EndExpansion
p_{k}(P_{k}-P_{0}),P^{\prime}-P_{0}\doteq%
%TCIMACRO{\tsum \nolimits_{k}}%
%BeginExpansion
{\textstyle\sum\nolimits_{k}}
%EndExpansion
p_{k}(P_{k}^{\prime}-P_{0}). \label{P-imbalance}%
\end{equation}
Let us first consider the exclusive approach. In equilibrium (with the
medium), we will have $P_{0}=%
%TCIMACRO{\tsum \nolimits_{k}}%
%BeginExpansion
{\textstyle\sum\nolimits_{k}}
%EndExpansion
p_{k\text{eq}}P_{k}=P$ ($\Delta P=0$). Since pressure fluctuations occur
\emph{even} in equilibrium \cite{Landau}, $\Delta P_{k}\doteq(P_{k}-P_{0}%
)\neq0,\forall k$. Thus, the pressure imbalance $\Delta P_{k}$\ is ubiquitous
and determines the microscopic internal work $d_{\text{i}}W_{k}\doteq\Delta
P_{k}dV=-d_{\text{i}}E_{k}$. We finally come to a very important observation
that $d_{\text{i}}E_{k}\neq0,\forall k$ (but must satisfy $d_{\text{i}}E=0$
\cite{Note}). The same discussion also applies to the primed SI quantities in
the inclusive approach so that%
\begin{equation}%
\begin{tabular}
[c]{l}%
$d_{\text{e}}W_{k}=-d_{\text{e}}E_{k}=P_{0}dV,d_{\text{e}}W_{k}^{\prime
}=-d_{\text{e}}E_{k}^{\prime}=-VdP_{0},$\\
$\ \ \ \ \ \ d_{\text{i}}^{\prime}W_{k}\doteq d_{\text{i}}W_{k},d_{\text{i}%
}W_{k}=-d_{\text{i}}E_{k},d_{\text{i}}^{\prime}W_{k}=-d_{\text{i}}%
E_{k}^{\prime}.$%
\end{tabular}
\ \ \ \ \ \ \label{Exclusive-Inclusive}%
\end{equation}
As discussed in \cite{Gujrati-Entropy2,Gujrati-GeneralizedWork} for
irreversibility, $d_{\text{i}}W_{k}$ is not necessarily nonnegative but the
IEA $d_{\text{i}}W$ is. \ \ \ \ \ \ \ \ \ \ \ \ \ \ \ \ \ \ \ \ \ \ \ 

\textsc{Jarzynski}$^{\prime}$\textsc{s Inclusive Approach} To make connection
with Jarzynski, we consider a simple mechanical system, a polymer chain, which
is being pulled by a force applied at one end; the other end is tethered so it
does not move. We consider an equilibrium ensemble of many single-chains in
state $\mathsf{A}$\textsf{ }at temperature $T_{0}$; the corresponding
microstate probabilities are given by $\left\{  p_{k0}\right\}  $. We focus on
one such chain but suppress the index $k$ for simplicity. The chain acts like
a spring with some SI spring potential $E(x)$, the \emph{exclusive energy},
where $x$ is the extension of the "spring" with respect to its mechanical
equilibrium position, where $E(x=0)=0$.\ Let $F_{0}$ be the\ external force,
which generates a displacement (elongation) $dx$ so that $dE=-Fdx$ for the
exclusive system, where $F=-\partial E/\partial x$ is the restoring spring
force, and we take $F$ and $F_{0}$ to point in the same direction. Writing
$dE=F_{0}dx-(F+F_{0})dx$, we can identify $dR=d_{\text{e}}E=F_{0}dx$ and
$d_{\text{i}}E=-F_{\text{t}}dx$, where $F_{\text{t}}=F+F_{0}$ is the net force
acting on the chain. The \emph{inclusive SI energy} $E^{\prime}$ is defined as
$E^{\prime}(x,F_{0})\doteq E(x)-F_{0}x$ in terms of the exclusive energy
$E(x)$ \cite{Note3}; dependence on other variables in $E$ and $E^{\prime}$ is
suppressed for simplicity as they are not relevant for our argument. The net
force $F_{\text{t}}$ must only vanish for mechanical equilibrium. As $\partial
E^{\prime}/\partial x=-F_{\text{t}}$, $E^{\prime}=E^{\prime}(x,F_{0})$ is a
function of $x$ and $F_{0}$ unless $F_{\text{t}}$ vanishes, \textit{i.e.},
$d_{\text{i}}E=0$. However, as said above regarding the pressure imbalance,
$d_{\text{i}}E\neq0$ even in the equilibrium ensemble. Thus, whenever there is
a mechanical force imbalance, $E^{\prime}(x,F_{0})$ is a function of two
variables, each of which plays the role of a work variable in the inclusive
approach so that the work in the inclusive approach in general is $dW^{\prime
}=-dE^{\prime}=xdF_{0}+F_{\text{t}}dx$ \cite{Note2} in which $dR^{\prime
}=d_{\text{e}}E^{\prime}=-xdF_{0}$. We must now average $dW^{\prime}$ to
obtain the thermodynamic average work $\left\langle dW^{\prime}\right\rangle
$. This will give rise to the components $d_{\text{e}}W^{\prime}=\left\langle
x\right\rangle dF_{0}$ and $d_{\text{i}}W^{\prime}=\left\langle F_{\text{t}%
}\right\rangle dx\geq0$. However, Jarzynski only treats $E^{\prime}$ as a
function of $F_{0}$ but not of $x$. This requires $\partial E^{\prime
}/\partial x=0$, \textit{i.e.,} $F_{\text{t}}=0$ or $d_{\text{i}}E^{\prime}%
=0$. (Even with this approximation, there has been some dispute in the
literature about the meaning of work
\cite{Rubi,Jarzynski-Rubi,Rubi-Jarzynski,Peliti-Rubi,Rubi-Peliti}, which is
simply a dispute between $dR$ and $dR^{\prime}$. We believe that both sides
are correct.) As is well known in nonequilibrium thermodynamics
\cite{Prigogine}, see also \cite{Cohen,Gujrati-GeneralizedWork}, it is the
$F_{\text{t}}$ term that results in the thermodynamic force \cite{Note2},
which then determines the dissipated work $d_{\text{i}}W^{\prime}=d_{\text{i}%
}W$. We have thus established that by neglecting this term, Jarzynski is
effectively considering a \emph{reversible process} so the JE does not capture
any irreversibility. A time-dependent protocol under certain conditions will
result in a force imbalance as the system is not able to respond to the
external stimuli to maintain $F_{\text{t}}=0$. Just having a time-dependent
protocol and not accounting for a nonzero $F_{\text{t}}$ microscopically is
not sufficient for irreversibility.

\textsc{Mystery Behind }$\left\langle R^{\prime}\right\rangle _{0}\geq\Delta
F^{\prime}$ \ \ Despite this, the use of Jensen's inequality gives rise to
$\left\langle R\right\rangle _{0}\geq\Delta F,\left\langle R^{\prime
}\right\rangle _{0}\geq\Delta F^{\prime}$, whose significance we must explain.
For this, we perform an explicit calculation in both approaches, for which, as
recently pointed out \cite{Gujrati-GeneralizedWork}, analogs of the JE are
available as \emph{identities}:%
\begin{equation}%
%TCIMACRO{\tsum \limits_{k}}%
%BeginExpansion
{\textstyle\sum\limits_{k}}
%EndExpansion
p_{k0}e^{\beta_{0}\Delta W_{k}}=e^{-\beta_{0}\Delta F},%
%TCIMACRO{\tsum \limits_{k}}%
%BeginExpansion
{\textstyle\sum\limits_{k}}
%EndExpansion
p_{k0}e^{\beta_{0}\Delta W_{k}^{\prime}}=e^{-\beta_{0}\Delta F^{\prime}},
\label{JarzynskiRelation}%
\end{equation}
by replacing $R_{k}$ in Eq. (\ref{JarzynskiRelation0}) by\ SI\ quantities
$-\Delta W_{k}$ and $-\Delta W_{k}^{\prime}$, respectively. Applying the
Jensen inequality results in $\left\langle -\Delta W\right\rangle _{0}%
\geq\Delta F$ and $\left\langle -\Delta W^{\prime}\right\rangle _{0}\geq\Delta
F^{\prime}$, respectively, which must be distinguished from the dissipation
inequalities $\left\langle -\Delta_{\text{e}}W\right\rangle =\left\langle
R\right\rangle \geq\Delta F$ and $\left\langle -\Delta_{\text{e}}W^{\prime
}\right\rangle =\left\langle R^{\prime}\right\rangle \geq\Delta F^{\prime}$,
respectively. As the distinction between $R_{k}$ and $\Delta E_{k}%
\doteq-\Delta W_{k}$ is due to irreversibility contribution, we will consider
a reversible isothermal process (no dissipation) so that $R_{k}=-\Delta W_{k}$
and $R_{k}^{\prime}=-\Delta W_{k}^{\prime}$. For the calculation, we consider
a simple example in which an ideal gas in a $1$-dimensional box of length $L$
expands quasistatically from $L_{\mathsf{A}}$ to $L_{\mathsf{B}}$; we let
$x\doteq L_{\mathsf{A}}/L_{\mathsf{B}}$ between \textsf{A }and \textsf{B}. As
there are no interparticle interactions, we can treat each particle by itself.
The microstates in the exclusive approach are those of a particle in the box
with energies determined by an integer $k:E_{k}=\alpha(k/L)^{2},\alpha=\pi
^{2}\hslash^{2}/2m$. Let $\beta_{0}$ denote the inverse temperature of the
heat bath. The gas remains in equilibrium at all times and $R_{\text{diss}}%
=0$. The partition function at any $x$ is given by
\[
Z(\beta_{0},L)=%
%TCIMACRO{\tsum \nolimits_{n}}%
%BeginExpansion
{\textstyle\sum\nolimits_{n}}
%EndExpansion
\exp(-\beta_{0}(n/L)^{2}\approx\sqrt{L^{2}\pi/4\alpha\beta_{0}}%
\]
for any $L\in\lbrack L_{\mathsf{A}},L_{\mathsf{B}}]$; in the last equation, we
have made the standard integral approximation for the sum. We then have%
\[
\beta_{0}F=-(1/2)\ln(L^{2}\pi/4\alpha\beta_{0}),E=1/2\beta_{0}.
\]

We can now compute the two work averages with $R_{k}=E_{k}(L_{\mathsf{B}%
})-E_{k}(L_{\mathsf{A}})$. For the Jarzynski average, we have%
\begin{equation}
\left\langle R\right\rangle _{0}=%
%TCIMACRO{\tsum \nolimits_{k}}%
%BeginExpansion
{\textstyle\sum\nolimits_{k}}
%EndExpansion
p_{k0}[E_{k}(L_{\mathsf{B}})-E_{k}(L_{\mathsf{A}})]=(x^{2}-1)/2\beta_{0},
\label{JarzynskiWork}%
\end{equation}
where we have used $E_{k}(L_{\mathsf{B}})-E_{k}(L_{\mathsf{A}})=(x^{2}%
-1)E_{k}(L_{\mathsf{A}})$. For the thermodynamic average, we use
$dE_{k}=-2E_{k}dL/L$ in Eqs. (\ref{ThermodynamicAverageWork}) or
(\ref{Thermodynamic Average}) to obtain%
\begin{equation}
\left\langle R\right\rangle =1//\beta_{0}\ln x=\Delta F.
\label{ThermodynamicWork}%
\end{equation}

It should be clear that it is the thermodynamic average work $\left\langle
R\right\rangle $ that satisfies the condition of equilibrium and not
$\left\langle R\right\rangle _{0}$, which is evidently different from
$\left\langle R\right\rangle $. Applying the Jensen inequality to the first
equation in Eq. (\ref{JarzynskiRelation}) with $\left\langle \bullet
\right\rangle _{0}$ for \textsf{E}, we obtain%
\[
e^{-\beta_{0}\left\langle R\right\rangle _{0}}\leq e^{-\beta_{0}\Delta F},
\]
yielding $\left\langle R\right\rangle _{0}\geq\Delta F$ and not $\left\langle
R\right\rangle \geq\Delta F$ as concluded by Jarzynski \cite{Jarzynski}. Let
us evaluate $\left\langle R\right\rangle _{0}-\Delta F$. Introducing
$y=1-x^{2}\geq0$ for expansion, we have%
\[
\left\langle R\right\rangle _{0}-\Delta F=[\ln(1-y)-y]//2\beta_{0}>0.
\]
The Jensen inequality is satisfied as expected, but the above
\emph{nonnegative difference} $\left\langle R\right\rangle _{0}-\Delta F$
makes no statement about any dissipation in the system, which is absent.

We now turn to the inclusive approach for which we need to determine the
equilibrium pressure $P_{0}$. We can determine it at any $L$ by its definition
given above Eq. (\ref{P-imbalance})~with $P_{k}=-\partial E_{k}/\partial
L=2E_{k}/L$. It is the same sort of calculation done above, and the result is
$P_{0}L=2E=1/\beta_{0}$. Therefore, $E^{\prime}-E=1/\beta_{0}$, a constant for
the process. As energy has just shifted by a constant, the physics is no
different from that in the exclusive approach. We find that between \textsf{A
}and \textsf{B}, $\Delta E_{k}^{\prime}=\Delta E_{k}$, so $\left\langle
R^{\prime}\right\rangle =\left\langle R\right\rangle =\Delta F=\Delta
F^{\prime}$, and $\left\langle R^{\prime}\right\rangle _{0}=\left\langle
R\right\rangle _{0}>\Delta F=\Delta F^{\prime}$. Incidentally, we also note
that $\Delta P_{k}\neq0,\forall k$ in this reversible expansion, even though
the corresponding thermodynamic force $\left\langle \Delta P\right\rangle =0$.

In summary, we have shown that the application of the Jensen inequality to Eq.
(\ref{JarzynskiRelation0}) or (\ref{JarzynskiRelation}) does not at all make
any statement about the second law so that $\left\langle R\right\rangle
_{0}>\Delta F$ or $\left\langle R^{\prime}\right\rangle _{0}>\Delta F^{\prime
}$ should not be confused with some generalized second law statement. We
further find that $\left\langle R\right\rangle _{0}\neq\left\langle
R\right\rangle $ and $\left\langle R^{\prime}\right\rangle _{0}\neq
\left\langle R^{\prime}\right\rangle $. It seems quite clear from our analysis
that the JE is based on the assumption$\ d_{\text{i}}^{\prime}W_{k}\doteq
d_{\text{i}}W_{k}=-d_{\text{i}}E_{k}=-d_{\text{i}}E_{k}^{\prime}=0$ (even
though its presence is ubiquitous microscopically) so it cannot capture any
irreversibility even though the work protocol is time dependent. On the other
hand, Eq. (\ref{JarzynskiRelation}) proposed by us captures irreversibility by
including $d_{\text{i}}E_{k}=d_{\text{i}}E_{k}^{\prime}$ but has limited
applicability unless we can determine $d_{\text{i}}E_{k}=d_{\text{i}}%
E_{k}^{\prime}$ in a real process, which seems very hard.


\begin{thebibliography}{99}
\bibitem {Note}Below is a brief introduction to the notation and the
significance of various modern terminology \cite{deGroot,Prigogine} for
readers who are unfamiliar with them. As usual, $\Sigma$ and $\widetilde
{\Sigma}$ form an isolated system $\Sigma_{0}$. Extensive quantities
associated with $\widetilde{\Sigma}$ and $\Sigma_{0}$\ carry a tilde
$\widetilde{\square}$ and a suffix $0$, respectively. As $\widetilde{\Sigma}$
is very large compared to $\Sigma$ and is in equilibrium, all its conjugate
fields $T_{0},P_{0},$ etc. carry a suffix $0$ as they are the same as for
$\Sigma_{0}$, and there is no irreversibility in $\widetilde{\Sigma}$. The
irreversibility is always ascribed to the system $\Sigma$, and is caused by
processes such as dissipation that are internal to the system. Quantities
without any suffix refer to the system. Any extensive system-intrinsic (SI)
quantity \cite{Note-SI} $q(t)$ of $\Sigma$ can undergo two distinct kinds of
changes in time: $d_{\text{e}}q(t)$ due to the exchange (suffix e) with the
medium and $d_{\text{i}}q(t)$ due to internal (suffix i) processes, and
$dq(t)\equiv d_{\text{e}}q(t)+d_{\text{i}}q(t)$; here $dq(t)\doteq
q(t+dt)-q(t),d_{\text{e}}q(t)\doteq-d\widetilde{q}(t)=-d_{\text{e}}%
\widetilde{q}(t)$ since $d_{\text{i}}\widetilde{q}(t)=0,$ and $d_{\text{i}%
}q(t)\doteq dq(t)+d\widetilde{q}(t)\equiv dq_{0}(t)$. In a reversible change,
$d_{\text{i}}q(t)\equiv0$. For example, the entropy change $dS\equiv
d_{\text{e}}S+d_{\text{i}}S$ for $\Sigma$; here, $d_{\text{e}}S=-d_{\text{e}%
}\widetilde{S}$ is the entropy exchange with the medium and $d_{\text{i}}S$ is
irreversible entropy generation due to internal processes within $\Sigma$; the
latter is also the entropy change $dS_{0}$\ of $\Sigma_{0}$. Similarly, if
$dW$ and $dQ$ represent the generalized work done by and the generalized heat
change of the system, then $dW\equiv d_{\text{e}}W+d_{\text{i}}W,$ $dQ\equiv
d_{\text{e}}Q+d_{\text{i}}Q$. Here, $d_{\text{e}}W$ and $d_{\text{e}}Q$ are
the work exchange and heat exchange with the medium, respectively, and
$d_{\text{i}}W\equiv$ $d_{\text{i}}W_{0}$ and $d_{\text{i}}Q\equiv$
$d_{\text{i}}Q_{0}$ are irreversible work done and heat generation due to
internal processes in $\Sigma$. For an isolated system such as $\Sigma_{0}$,
the exchange quantity vanishes so that $dq_{0}(t)=d_{\text{i}}q_{0}(t)$. The
first law in terms of generalized quantities of the system is given by
$dE=dQ-dW=TdS-dW$, but in terms of exchange quantities is given by
$dE=d_{\text{e}}Q-d_{\text{e}}W$. This requires the identity $d_{\text{i}%
}Q\equiv d_{\text{i}}W$ \cite{Gujrati-II,Gujrati-Entropy2}. We denote the
accumulation of $d_{\text{e}}W,dW$, etc. in a process by $\Delta_{\text{e}%
}W,\Delta W$, etc. As $E$ can only be changed by exchange and not by any
internal processes, $dE=d_{\text{e}}E,d_{\text{i}}E=0$. These partitions for
various $dq$'s are easily extended at the microscopic level to microstates as
shown in \cite{Gujrati-Stat}, with some noticeable differences. For example,
we will show here that $d_{\text{i}}E_{k}\neq0$ for the $k$th microstate
ubiquitously even in a reversible process, even though the $d_{\text{i}%
}E=d_{\text{i}}Q-d_{\text{i}}W=0$, which is nothing but the first law for
internal processes.
\bibitem {Note-SI}Any extensive or intensive quantity such as the energy $E$,
entropy $S$, volume $V$, temperature $T$, restoring force $F$, generalized
work $\Delta W$ and heat $\Delta Q$, etc. that depend on the system alone are
called system-intensive (SI) quantities. The external temperature $T_{0}$,
pressure $P_{0},$ force $F_{0}$, etc. are not SI quantities but can control
the exchange quantities $d_{\text{e}}q$ for the system and determine the
thermodynamic forces \cite{Note2} such as $T-T_{0},P-P_{0},F+F_{0}$, etc.
which control the system's approach to equilibrium.
\bibitem {Note-R}In terms of our notation introduced in \cite{Note},
$\left\langle R\right\rangle =-\Delta_{\text{e}}W$ and differs from $-\Delta
W$ due to the presence of irreversibility $\Delta_{\text{i}}W$.
\bibitem {Note-JE-GeneralizedWork}A careful analysis
\cite{Gujrati-GeneralizedWork}\ shows that the correct identification is in
terms of the SI generalized work $dW_{k}=-dE_{k}$ over the entire process
$\mathcal{P}_{0}$ with $d_{\text{i}}E_{k}\neq0$ in general. When $d_{\text{i}%
}E_{k}\neq0$, we need to recall that $dR_{k}=d_{\text{e}}E_{k}$. Being an
exchange quantity, it is not an SI\ quantity \cite{Note-SI}. In that case, Eq.
(\ref{JarzynskiRelation0}) will relate a non-SI quantity with an SI quantity,
which casts doubt on its validity.
\bibitem {InternalEQ}We assume the system to be in internal equilibrium
\cite{Gujrati-II,Gujrati-Entropy2} so that $E(t)$ is a state function of
$S(t)$ and $V(t)$. It is possible to go beyond the internal equilibrium
hypothesis and allow an explicit time dependence in $E$ but we will not do
this here.
\bibitem {Note-Legendre}A Legendre transform of $f(x,y)$ with respect to $y$
is given by $f^{\prime}\left(  x,Y\right)  \doteq f(x,y)-yY$, where
$Y\doteq\left(  \partial f/\partial y\right)  $. Thus, $\left(  \partial
f^{\prime}/\partial y\right)  =0$. Conversely, for $f^{\prime}$ to be
independent of $y$, we have $\left(  \partial f^{\prime}/\partial y\right)
=0$ from which it follows that $Y=\left(  \partial f/\partial y\right)  $. For
the nonequilibrium enthalpy, $P_{0}$ has nothing to do with the derivative
$\partial H/\partial V=-\Delta P$ except when $\Delta P=0$.
\end{thebibliography}
\end{document}